\documentstyle[prl,aps,twocolumn,psfig]{revtex}
\def\Mag{m}
\draft
\begin{document}
\twocolumn[\hsize\textwidth\columnwidth\hsize
\csname @twocolumnfalse\endcsname
\title{Field induced ordering in highly frustrated antiferromagnets}
\author{M. E. Zhitomirsky,$^1$ A. Honecker,$^2$ and 
        O. A. Petrenko$^{3,4}$}
\address{$^1$Theoretische Physik, 
         ETH H{\"o}nggerberg, CH--8093 Z{\"u}rich, Switzerland}
\address{$^2$Institut f\"ur Theoretische Physik, TU Braunschweig,   
         Mendelssohnstr.\ 3, D--38106 Braunschweig, Germany} 
\address{$^3$ISIS Facility, Rutherford Appleton Laboratory, 
         Chilton, Didcot, OX11~0QX, United Kingdom}
\address{$^4$University of Warwick, Department of Physics, 
         Coventry, CV4~7AL, United Kingdom}
\date{\today}
\maketitle
\begin{abstract}
We predict that an external field can induce a spin order in
highly frustrated classical Heisenberg magnets. 
We find analytically stabilization of collinear states by 
thermal fluctuations at a one-third of the saturation field 
for kagome and garnet lattices and at a half of the saturation 
field for pyrochlore and frustrated square lattices. This effect 
is studied numerically for the frustrated square-lattice 
antiferromagnet by Monte Carlo simulations for classical spins and 
by exact diagonalization for $S=1/2$. The field induced
collinear states have a spin gap and produce magnetization plateaus.
\end{abstract}
]

Frustrated magnets, classical and quantum, exhibit spectacular 
and often unexpected behaviors \cite{frust}. A few of them
do not order at any $T>0$. Extensive residual entropy in 
the classical ground state in such cases prevents the realization
of the usual order by disorder scenario \cite{obdo}. 
An external magnetic field changes the degeneracy and topology
of the ground state manifold stabilizing, eventually,
the (nondegenerate) saturated state at $H>H_{\rm sat}$. 
If the order by disorder effect occurs in a finite field
and suppresses residual entropy, then, such an effect can be used for 
practical applications. During a demagnetization process a spin system 
has to regain its entropy and, therefore, the whole crystal will 
cool down. Examples of geometrically frustrated AFMs on kagome 
\cite{kagome}, pyrochlore \cite{pyro}, and garnet \cite{ggg} lattices 
include magnetic compounds with rather small exchange constants 
opening the way for experimental tests of their finite field behavior.
 
In this Letter, we predict that fluctuations stabilize 
collinear spin configurations at rational values of the field
$H/H_{\rm sat}=1/2$ or $1/3$.
We present an analytical proof of the field 
induced ordering driven by thermal fluctuations and suggest a similar 
role for quantum fluctuations on the basis of spin-wave and numerical 
results. In fact, an ordered spin phase in a finite field has already 
been observed in gadolinium garnet Gd$_3$Ga$_5$O$_{12}$, though dipolar 
anisotropy plays a crucial role in this material \cite{ggg}.

To be specific we use as an example the frustrated square-lattice 
antiferromagnet (FSAFM) \cite{singlet} in an external field:
\begin{equation} 
\hat{\cal H} = J \sum_{\rm n.n.} {\bf S}_i\cdot {\bf S}_j
    + J' \sum_{\rm n.n.n.} {\bf S}_i \cdot {\bf S}_j -H\sum_i S^z_i \ . 
\label{H}
\end{equation}
First, we present analytical arguments, making them as general 
as possible, in order to include the other highly 
frustrated AFMs. Second, we consider numerical results for 
classical and quantum FSAFMs in a magnetic field, which confirm 
our predictions.

\begin{figure}[ht]
\centerline{\psfig{figure=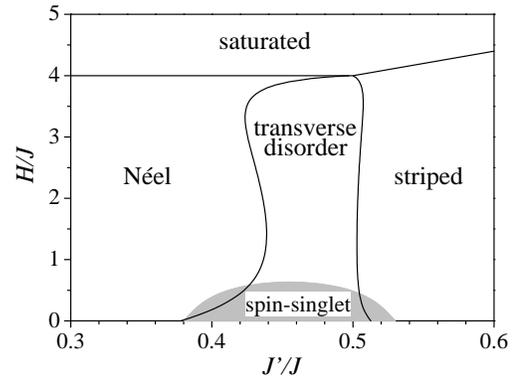,width=0.8\columnwidth}}
\caption{Zero temperature phase diagram of the spin-1/2 FSAFM 
in external field obtained from the linear spin-wave theory.
The shaded region indicates the stability region of the quantum 
disordered state with a spin gap.}
\label{diagram}
\end{figure}

Let us briefly discuss magnetically ordered phases of the FSAFM.
For small diagonal exchange $J'<0.5J$, classical spins form 
the N\'eel state in zero field. At $J'>0.5J$ the classical ground 
state consists of two $\sqrt{2}\!\times\!\sqrt{2}$ interpenetrating 
antiferromagnets, which are locked by fluctuations in a striped 
AFM state described by a single wave-vector $(\pi,0)$ or $(0,\pi)$ 
\cite{obdo}. 
An applied magnetic field cants the spins and creates 
an easy plane for the AFM sublattices.
Thus, both the N\'eel and striped states have a transverse 
spin order in a finite field.  Thermal and quantum fluctuations
destroy the transverse order in the vicinity of the highly frustrated 
point $J'=0.5J$. As an illustration we present in Fig.~\ref{diagram}
the phase diagram for the spin-1/2 model at $T=0$ obtained in 
the linear spin-wave theory.
This phase diagram agrees with the previous zero-field  
studies \cite{singlet}, which suggest a quantum 
disordered ground state for $S=\frac{1}{2}$ near $J'=0.5J$. 
Since the disordered singlet phase has a spin gap,
it becomes unstable above a finite critical field.

We now focus on the classical critical point $J'=0.5J$, where
the Hamiltonian (\ref{H}) can be written up to a constant term 
as a sum over edge-sharing plaquettes:
\begin{equation} 
\hat{\cal H}= \frac{1}{2^n} \sum_\alpha\left\{J \left|
{\bf L}_\alpha\right|^2 
- H L^z_\alpha \right\} ,
\label{Hpl}
\end{equation}
where ${\bf L}_\alpha = \sum_{i\in\alpha} {\bf S}_i$ is the total spin 
in plaquette $\alpha$ and $n=2$. The block form 
of the spin Hamiltonian (\ref{Hpl}) is common to all highly frustrated 
spin models. For AFMs on kagome and garnet lattices the blocks 
are triangles and for a pyrochlore AFM they are tetrahedra
with the corner-sharing arrangements and $n=1$ in all three cases. 

A spin configuration minimizes the energy (\ref{Hpl}) at $H=0$ 
provided ${\bf L}_\alpha=0$ for each plaquette. 
This constraint can be satisfied for many classical states. 
We estimate the number of continuous degrees of freedom in 
the ground state of the $N$-site FSAFM as $\sim N^{1/2}$. For pyrochlore 
and kagome AFMs the degeneracy is larger and the dimensionality of 
the ground state manifold scales with $N$ \cite{kagome,pyro,MoeCha}. 
Thus, zero-field properties are not universal for disordered
frustrated AFMs. We now show that a universal behavior does appear 
in a magnetic field.  

\begin{figure}[ht]
\centerline{\psfig{figure=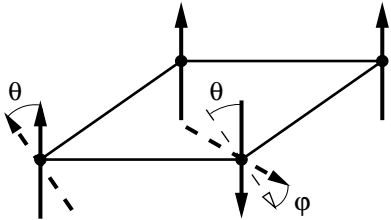,width=0.6\columnwidth}}
\caption{The collinear spin structure for a four-spin block 
at $H=0.5H_{\rm sat}$. Rotations by angles $\theta$ and $\varphi$ 
parameterize the zero-energy and soft modes in this state.}
\label{uuud}
\end{figure}

In a finite field the classical energy 
(\ref{Hpl}) is minimized for spin configurations with the plaquette 
magnetization $L^z_\alpha = H/(2J)$. There are many degenerate classical 
states which satisfy this constraint below
the saturation field $H_{\rm sat}=8JS$ (pyrochlore and FSAFM)
or $6JS$ (kagome and garnet). Generally, all these states are 
noncollinear as, e.g., canted N\'eel and striped phases for FSAFM.
Collinear spin arrangements
appear only at special rational values of the applied field:
at $H_c=\frac{1}{2}H_{\rm sat}$ for 4-spin blocks (FSAFM, pyrochlore),
the up-up-up-down (uuud) structure, see Fig.~\ref{uuud}, and 
at $H_c=\frac{1}{3}H_{\rm sat}$ for 3-spin blocks (kagome, garnet),
the up-up-down structure. The distribution of down-spins on the lattice 
in a collinear state is not unique: it only obeys the 
one down-spin per plaquette condition.

Zero-energy modes correspond to continuous distortions of a given 
classical ground state, which do not violate the magnetization 
constraint. To construct a zero mode for the collinear states
described above one has to draw an (open) line through the lattice points 
with adjacent sites on a line occupied by antiparallel
spins with no plaquettes crossed more than once. 
A zero mode consists of a simultaneous rotation
of all spins along the line by an angle $\theta$ (Fig.~\ref{uuud}).
The complete set of zero modes is constructed when a set of
such lines goes through all up-spins. 
Zero modes connect a uuud state to noncollinear ground state
configurations. Hence, the collinear states 
are singular points on the ground state manifold. There are no
other collinear states in the local neighborhood of a selected
collinear state.

Moessner and Chalker \cite{MoeCha} have recently discussed the order 
by disorder phenomenon in zero field for classical frustrated models in 
terms of a local topology of the ground state manifold. They 
reached definite conclusions only for an $XY$ magnet on the pyrochlore
lattice and for a Heisenberg kagome lattice AFM. We show that a
similar analysis predicts stabilization of collinear states near 
special rational values of the applied field.
Let $\bf x$ denote the 
coordinates on the ground state manifold and $\bf y$ be the transverse 
directions, which span the rest of the configuration space. Low-energy 
excited states are described by the quadratic Hamiltonian ${\cal H}_2 = 
\sum_l \epsilon_l({\bf x}) y_l^2$ resulting in a probability distribution
\begin{equation}
Z({\bf x}) = \int dy_l e^{-\beta {\cal H}_2} \propto
\prod_l[T/\epsilon_l({\bf x})]^{1/2}
\end{equation}
over the ground state manifold at low temperatures. At a special 
point ${\bf x}_0$ some of the stiffnesses $\epsilon_l({\bf x}_0)$ 
may vanish making $Z({\bf x}_0)$ divergent. The corresponding 
coordinates $y_l$ describe soft modes. The appearance of the order 
by disorder depends on a number of soft modes, which exist for a given
classical ground state ${\bf x}_0$. 

A special feature of the collinear states with finite magnetization 
is that each zero mode generates exactly one soft mode. 
They are constructed in the following way. 
First, all spins along a zero mode line are rotated
by $\theta$ about the axis perpendicular to the field
and, second, all down spins on the same line are rotated by an angle
$\varphi$ about the field direction, see Fig.~\ref{uuud}.
The total spin of each plaquette in a deformed state
is ${\bf L}_\alpha \approx 
S(0,\theta\varphi,2)$ for small $\theta$ and $\varphi$ 
and contributes to an energy increase 
$\sim \theta^2\varphi^2$. Thus, we identify $x_l$ with $\theta$
and $y_l$ with $\varphi$.

To allow the order by disorder selection, not only has the probability
density $Z({\bf x}_0)$ to diverge, but the statistical weight 
$\int Z({\bf x})d{\bf x}$ must be concentrated entirely
near the collinear spin states. We check this by integrating 
$Z({\bf x})$ in the $D$-dimensional neighborhood
of a collinear state parameterized by zero modes.
Since each zero mode has one soft mode 
with $\epsilon_s({\bf x})\sim x^2$, the integral
\begin{equation}
\int Z({\bf x}) d{\bf x} \sim \int x^{-D} d^D x
\label{integral}
\end{equation}
diverges independently of the actual value of $D$. 
There is a vanishingly small  probability of finding a spin 
system in one of the noncollinear states surrounding a given
collinear configuration. Hence, the order by disorder selection occurs: 
thermal fluctuations stabilize a discrete set of collinear states. This 
result relies only on the special block structure (\ref{Hpl}) 
of the spin Hamiltonian and the symmetry properties of the collinear states. 
Therefore, the order by disorder 
in an external field is a universal effect and occurs for 
all frustrated Heisenberg AFMs: apart from the FSAFM the uuud 
states at $H=\frac{1}{2}H_{\rm sat}$ are stabilized for a pyrochlore 
AFM and the uud states appear for Heisenberg AFMs on kagome and garnet 
lattices at $H=\frac{1}{3}H_{\rm sat}$. A subsequent selection 
between collinear states with different patterns of down-spins is made 
to maximize $D$ in Eq.~(\ref{integral}) and corresponds 
in the case of FSAFM to the most symmetric ${\bf q}=0$ uuud state.
The above proof of the field induced ordering
is based on the special property of the collinear states with 
finite magnetization. Namely, an equal number of zero and soft modes.
For a Heisenberg magnet on the pyrochlore lattice in zero field 
the same equality holds only approximately: in the leading order in the 
number of spins $N$. Therefore, no analytical
conclusion has been reached in a zero field case, while numerical
simulations indicated no ordering \cite{MoeCha}.

For quantum models in the limit $S\gg 1$ we check the relative stability 
of different states by comparing their zero-point oscillation energies: 
$\frac{1}{2}\sum_{\bf k}\omega_{\bf k}$. The magnon spectrum of 
the ${\bf q}=0$ uuud state has four modes:
\begin{equation}
 \omega_{1,2} = JS\sqrt{\eta_{\bf k}} \ , \ \ 
 \omega_{3,4} = 2JS\pm JS\sqrt{4-\eta_{\bf k}} 
\label{spec}
\end{equation}
with $\eta_{\bf k}= (1-\cos k_x)(1-\cos k_y)$. Zero-point contributions 
to the energies of the ${\bf q}=0$ uuud state, 
the N\'eel state, and the striped AFM  at $J'=0.5J$ and 
$H=\frac{1}{2}H_{\rm sat}$ are $0.57JS$, $0.69JS$, and $0.81JS$, 
respectively. (They have the same classical energy.) 
Thus, this comparison suggests that quantum fluctuations 
also select the collinear states in a magnetic field due to 
their large number of soft modes. 
Collinear states preserve the $O(2)$-rotational
symmetry. Hence, their renormalized magnon spectrum 
becomes gapped and plateaus arise on the magnetization curve. 
The spin-density wave of the ${\bf q}=0$ uuud state 
in FSAFM is a superposition of spin harmonics with wave-vectors
$(0,0)$, $(\pi,\pi)$, $(\pi,0)$ and $(0,\pi)$. 

\begin{figure}[ht]
\centerline{\psfig{figure=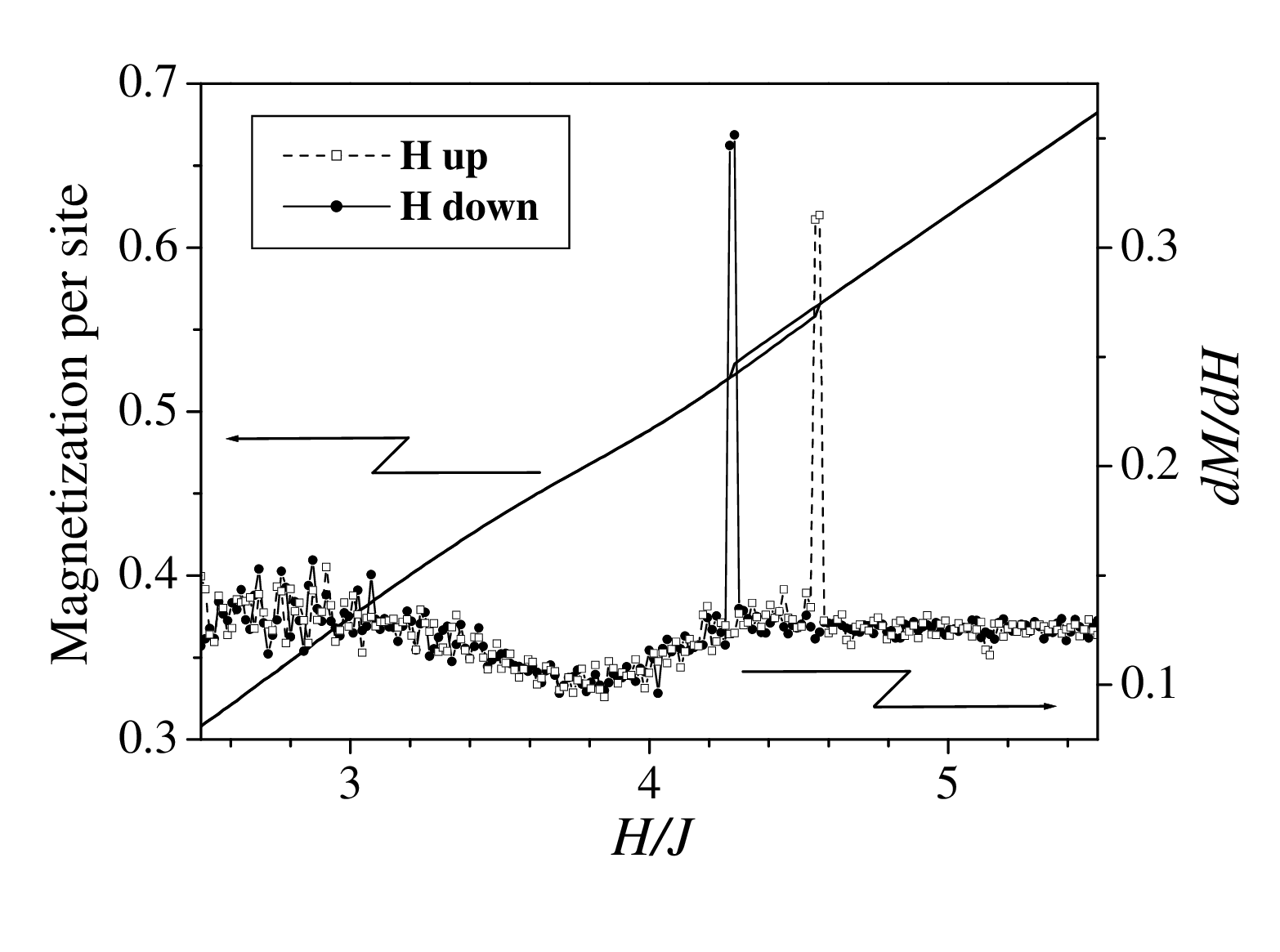,width=0.9\columnwidth}}
\vspace{2mm}
\centerline{\psfig{figure=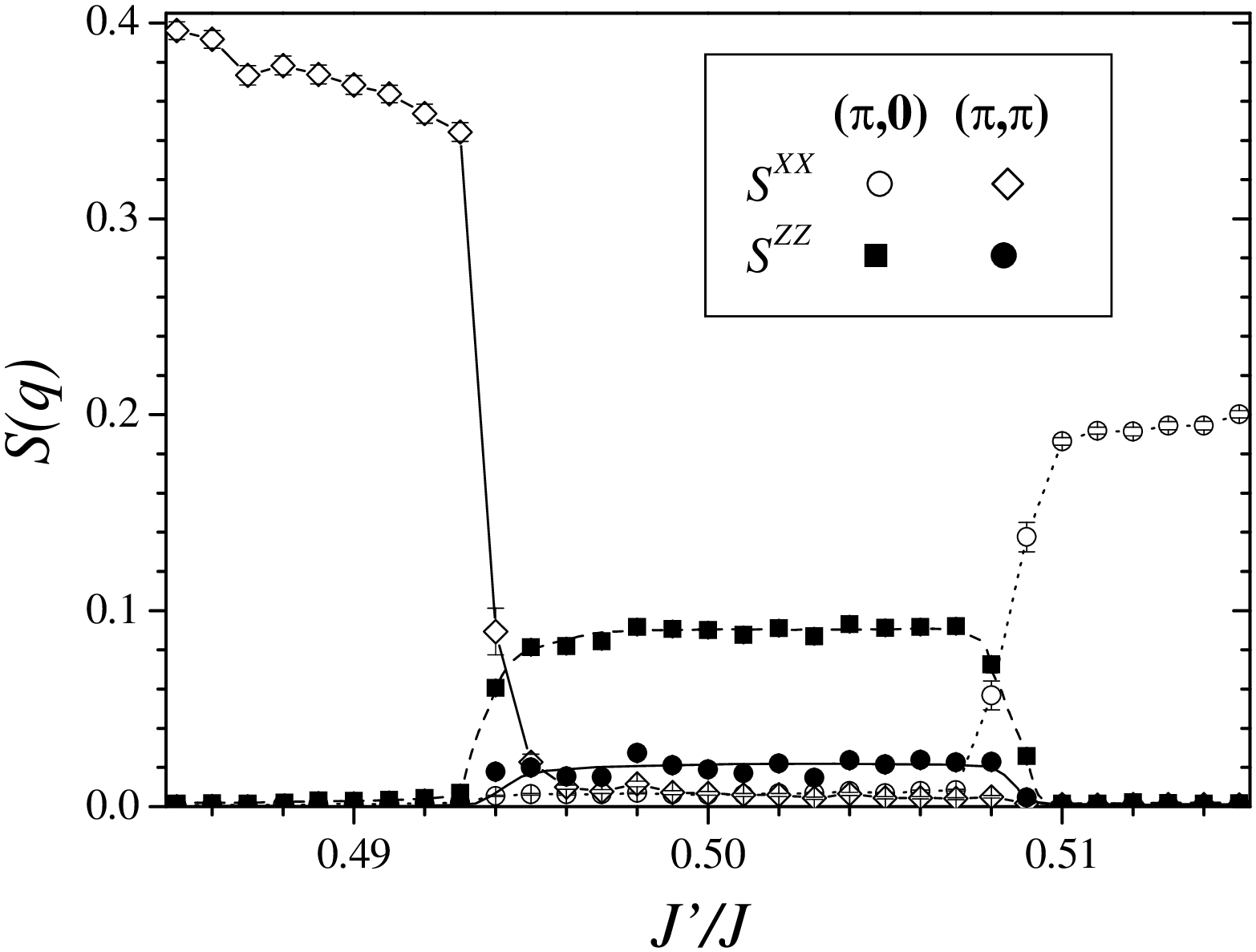,width=0.78\columnwidth}}
\caption{Monte Carlo results for the classical FSAFM: 
the magnetization curve for $J'/J=0.5$ (top panel) and
the static structure factor $S^{\alpha\beta}(q)$ vs.\ the 
frustration parameter for $H=4J$ (bottom panel) 
on a $24\!\times\!24$ lattice at $T=0.1J$.}
\label{MC}
\end{figure}

To test these predictions we have performed Monte Carlo (MC)
simulations for the model Eq.~(\ref{H}) with unit vector spins.
Fig.~\ref{MC}, top panel shows the magnetization curve at 
$T=0.1J$ obtained with $2\times 10^5$ MC steps per spin for each 
point. The field derivative of the magnetization has a dip 
around $H=4J=\case{1}{2}H_{\rm sat}$, which indicates the presence of 
a new plateau phase. A jump in the magnetization and a hysteresis
of about $\Delta H\simeq 0.3J$ suggests a first-order transition to a
high-field phase. To determine the nature of the spin state at the
plateau we calculated different components of the static structure
factor $S^{\alpha\beta}(q)=\frac{1}{N^2}\sum_{r,x}{\rm e}^{i{\bf
q}\cdot{\bf x}} \langle S^{\alpha}_r S^{\beta}_{r+x}\rangle$. The
N\'eel and the collinear states have nonzero transverse components
$S^{xx}=S^{yy}$ at ${\bf q}=(\pi,\pi)$ and ${\bf q}=(\pi,0)$ or
$(0,\pi)$, respectively. The uuud state has nonzero
longitudinal components $S^{zz}(q)$ at all the above vectors
simultaneously. The static structure factor presented in
Fig.~\ref{MC}, bottom panel was obtained by averaging 50 `instant
shots' separated by $10^3$ MC steps. In the region of weak (strong)
diagonal exchange $J'$ the data clearly support the N\'eel (striped)
type of spin correlations.  Nonzero harmonics in the longitudinal
structure factor both at ${\bf q}=(\pi,0)$ and ${\bf q}=(\pi,\pi)$
exist only for $0.494<J'/J<0.508$. We have also checked that neither
a lower field of $H=2J$, nor a higher field of $H=6J$ induces a
nonzero value of $S^{zz}(q)$ at these points in the Brillouin zone.
Thus, these results unambiguously identify the spin configuration on
the plateau as the ${\bf q}=0$ uuud state.

The uuud phase breaks the translational
symmetry in such a way that the wave-vectors belonging to different
irreducible representations of the space group are mixed: $(\pi,\pi)$
vs.\ $(\pi,0)$ and $(0,\pi)$. Hence, the spin structure of the uuud
phase is a mixture of different order parameters. In such a case the phase
transition to a disordered spin liquid state at the high-field
end of the plateau, Fig.~\ref{MC}, must go either via a
first order transition or in several steps with an intermediate
supersolid state.

\begin{figure}[ht]
\centerline{\psfig{file=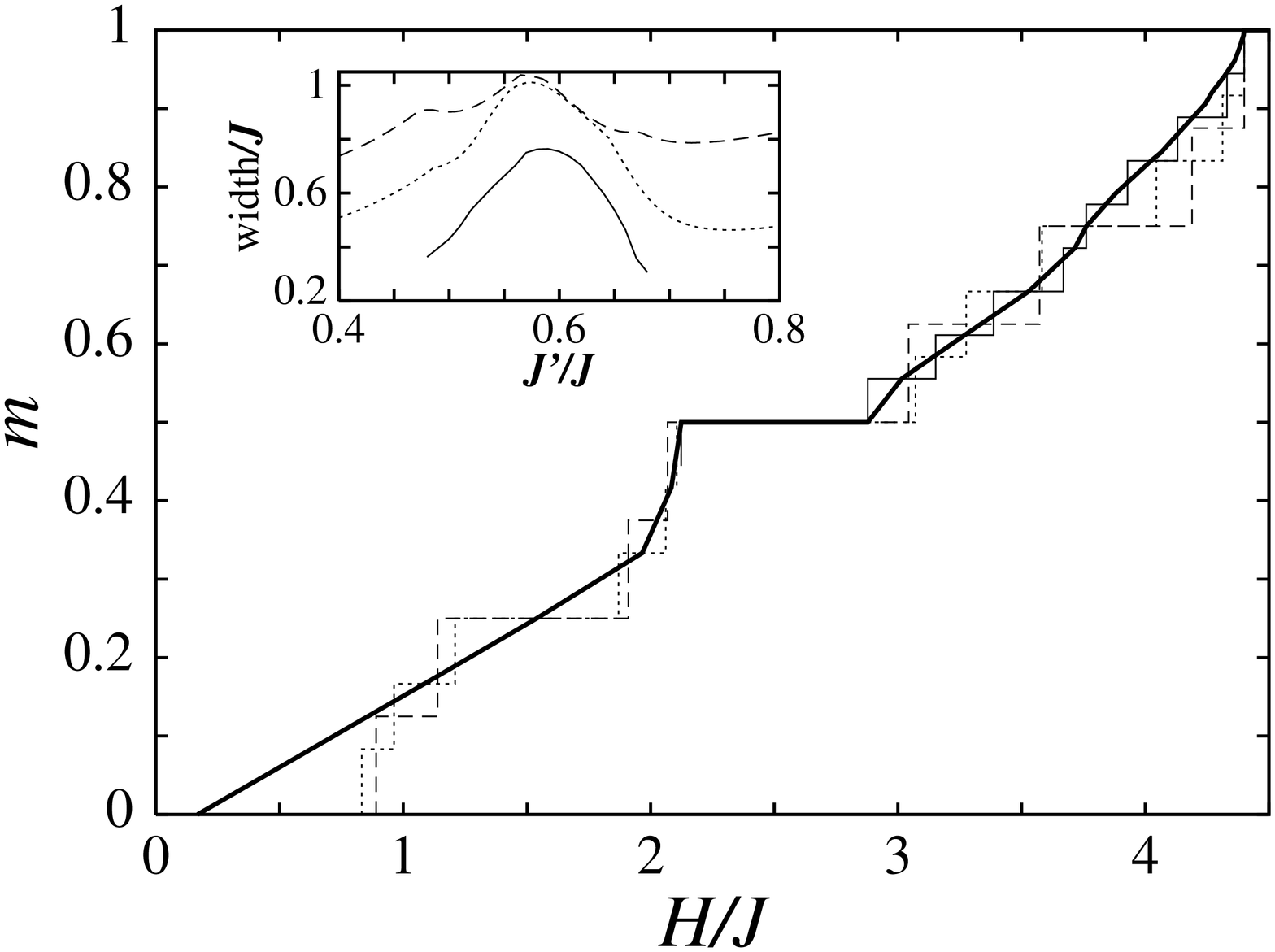,width=0.8\columnwidth,angle=270}}
\vspace{1mm}
\centerline{\psfig{file=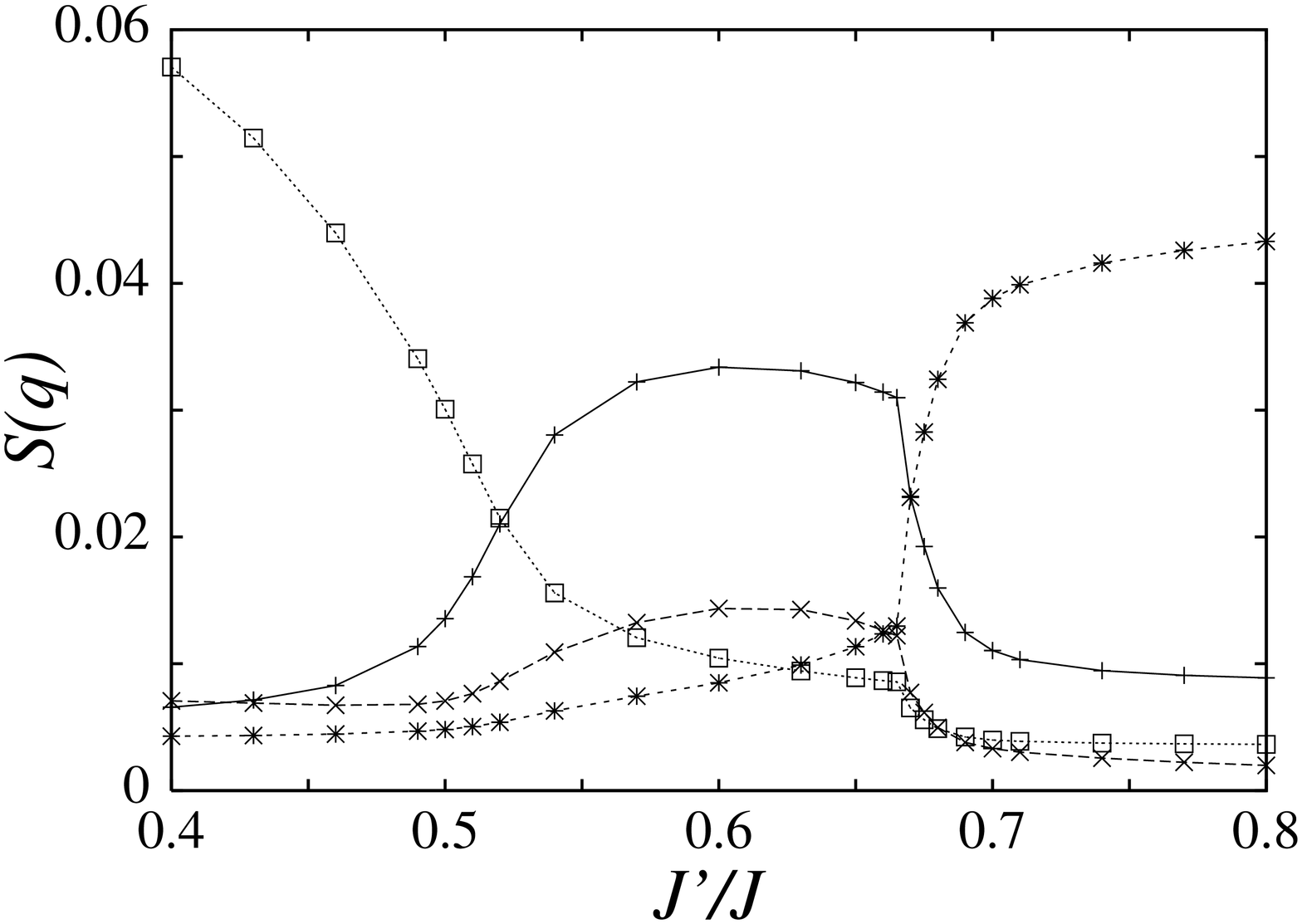,width=0.83\columnwidth,angle=270}}
\caption{Exact diagonalization results for the spin-1/2 FSAFM.
Top panel: the magnetization curve for $J'/J = 0.6$.
Inset: width of the finite-size plateau with $\Mag=\frac{1}{2}$ as
a function of $J'/J$. The lines are for lattices of size $4 \times 4$ 
(dashed), $6 \times 4$ (dotted), and $6 \times 6$ (full). The bold line 
is an extrapolated curve. Bottom panel: static structure factor as a 
function of $J'/J$ at $\Mag = \frac{1}{2}$ on a $6 \times 6$ cluster. 
$S^{zz}(0,\pi)$ is shown by `$+$', $S^{zz}(\pi,\pi)$ by 
`$\times$', $S^{xx}(0,\pi)$ by `$\rlap{+}\times$', and $S^{xx}(\pi,\pi)$ 
by squares.}
\label{ED}
\end{figure}

We have also studied the quantum spin-1/2 model (\ref{H}) at $T=0$ by
Lanczos diagonalizations of finite clusters. Fig.~\ref{ED}
presents the magnetization $m=M/S$ vs.\ field at $J'/J = 0.6$. There
are no magnetization plateaus in the thermodynamic limit for
$\Mag>1/2$. To determine whether the plateau at $\Mag=\frac{1}{2}$
remains after finite size scaling we show its width in the inset for
three cluster sizes as a function of $J'/J$.  If a plateau disappears
in the thermodynamic limit, its width should decrease as $\Delta
m=1/N$. Therefore, the ratio of the plateau-widths of the
$4\!\times\!4$ and $6\!\times\!6$ clusters should be $16/36$ or
less if the magnetization curve has a nonzero slope for
$N\rightarrow\infty$. The ratio $16/36$ is exceeded for $\Mag=1/2$
only in the interval $0.49\lesssim J'/J\lesssim 0.66$, where we
would expect to have a plateau on an infinite lattice.

Quantum fluctuations have a strong effect on the stability of the
$\Mag=\frac{1}{2}$ plateau. Both, the width of the plateau for the
spin-1/2 system and the range where it appears are larger than for
the classical model. In addition, the parameter range for the plateau
is shifted asymmetrically around the classical critical point
$J'=0.5J$. The plateau is most pronounced for $J' \approx 0.6J$.
This is the value we used for the presented magnetization curve. The bold line
in Fig.~\ref{ED} was obtained by connecting the mid-points of the
steps for the largest available systems size, except for the plateau
at $\Mag = \frac{1}{2}$, for which the corners of the $6\!\times\!6$
cluster data were used. The value of the spin gap ({\it i.e.}\ the
boundary of the $\Mag=0$ plateau) was taken from Kotov {\it et
al\/}.~\cite{singlet}. Fig.~\ref{ED} also shows the peaks of the
static structure factor vs.\ $J'/J$ for the $\Mag=\frac{1}{2}$ plateau. 
The peaks in $S^{zz}(q)$, which indicate the presence of the uuud state, 
exist for $0.51\lesssim J'/J \lesssim 0.67$.

The discussed mechanism for the magnetization plateaus in highly frustrated
AFMs has to be contrasted with that of weakly coupled 1D spin systems 
(see, e.g., \cite{1D}). There, magnetization plateaus correspond 
to disordered states with a gap, 
that are stable under a weak higher-dimensional coupling, while
intermediate gapless regions are immediately ordered at sufficiently
low temperatures once a higher-dimensional coupling is switched on.
Here, we found that an external field induces a long-range collinear
ordering with an excitation gap on the plateau, whereas intermediate 
gapless regions remain disordered.
Further exploration of such possibilities should also be
important for the general problem of classifying magnetization
plateaus in two dimensions \cite{2D}.

We thank J.T. Chalker, T.M. Rice, and O.P. Sushkov for helpful discussions. 
A.H. is grateful to
the Institut f\"ur Theoretische Physik of the ETH Z\"urich for
hospitality during the course of this project, to the Alexander von
Humboldt-foundation for financial support and to the
Max-Planck-Institut f\"ur Mathematik, Bonn for allocation of CPU time.
The work of M.E.Z. was supported by the Swiss National Fund.

\end{document}